\documentclass[aps,prl,twocolumn,showpacs,amsmath,amssymb,superscriptaddress,floatfix]{revtex4}

\usepackage{graphicx,color}

\usepackage{bm}  
\usepackage{graphicx} 
\usepackage{amssymb}
\usepackage{amsmath}
\usepackage{dcolumn} 

\usepackage[hypertex]{hyperref}

\begin{document}

\title{
Lattice model of three-dimensional 
topological
singlet superconductor
with time-reversal symmetry
}

\author{Andreas P.\ Schnyder}
\affiliation{Kavli Institute for Theoretical Physics,
  University of California,
  Santa Barbara,
  CA 93106,
  USA}

\author{Shinsei Ryu}
\affiliation{
Department of Physics, 
  University of California, Berkeley, CA 94720, USA
}

\author{Andreas W.\ W.\ Ludwig}
\affiliation{Department of
             Physics, University of California,
             Santa Barbara, CA 93106, USA}

\date{\today}

\begin{abstract}
We study topological phases of time-reversal invariant singlet superconductors in
three spatial dimensions.
In these particle-hole symmetric systems the topological phases are characterized by an 
even-numbered winding number $\nu$.
At a two-dimensional (2D) surface
the topological properties of this quantum state manifest themselves through
the presence of $\nu$ flavors of gapless Dirac fermion surface states, 
which are robust against localization from random impurities.
We construct a tight-binding model on the diamond lattice  that 
realizes  a topologically nontrivial phase,
in which the  winding number takes the value $\nu =\pm 2$.
Disorder corresponds to a (non-localizing) random SU(2) gauge potential
for the surface Dirac fermions, 
leading to a power-law density of states $\rho(\epsilon) \sim \epsilon^{1/7}$.
The bulk effective field theory is 
proposed to be the (3+1) dimensional SU(2) Yang-Mills 
theory with a theta-term at $\theta=\pi$.
\end{abstract}

\pacs{73.43.-f, 73.20.At, 74.25.Fy, 73.20.Fz, 03.65.Vf}

\maketitle

Bloch-Wilson band insulators are commonly believed to be simple and well understood electronic states of matter.
However, recent theoretical~\cite{KaneMele, bernevig06,moore07,Fu06_3Da,Fu06_3Db,Schnyder08} 
and experimental~\cite{konig07,hasan07} progress has shown that band insulators can exhibit 
unusual and conducting boundary modes, which are topologically protected, 
analogous to the edge states of the integer quantum Hall effect (QHE). 
These so-called $\mathbb{Z}_2$ topological insulators 
(also known as `quantum spin Hall' insulators),
which exist in two- and three-dimensional (3D) \emph{time-reversal invariant} (TRI) systems, are characterized by
a topological invariant, similar to the Chern number of the integer QHE.
Given these newly discovered topological states, one might wonder whether  
there exists a general organizing principle for topological insulators. 
Indeed, the integer QHE and the $\mathbb{Z}_2$ topological insulators are in fact part of 
a larger scheme,
discussed  in 
Ref.~\cite{Schnyder08},
which provides an exhaustive classification of topological insulators 
and superconductors in terms of spatial dimension and 
the presence or absence of 
the two most generic symmetries of the Hamiltonian, time-reversal and particle-hole
symmetry \cite{Kitaev08}.
\begin{figure}
  \begin{center}
(a)  \includegraphics[width=0.16\textwidth,clip]{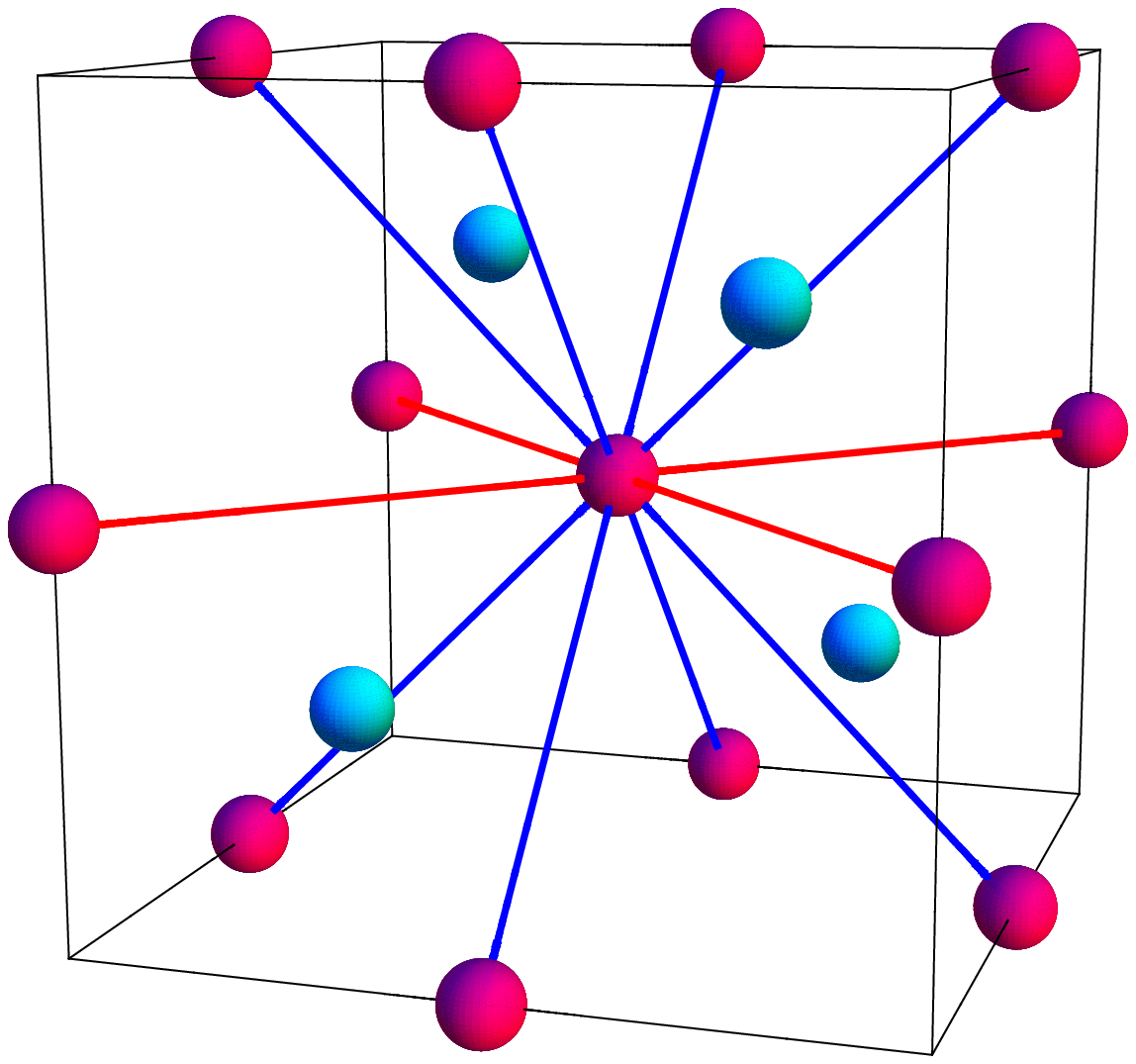} 
(b) \includegraphics[width=0.16\textwidth,clip]{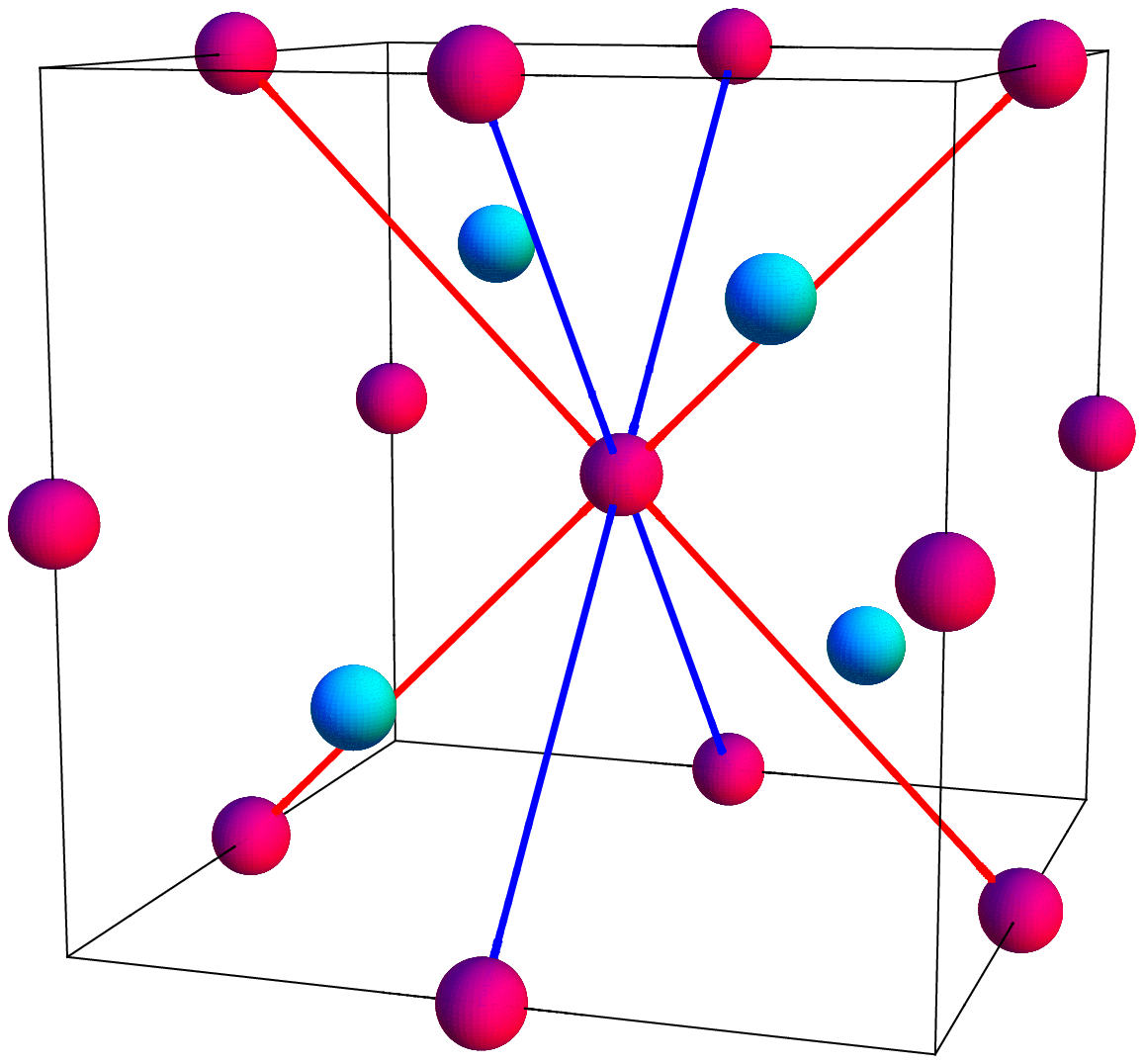} 
\caption{
\label{fig: diamond lattice}
(Color online)
The diamond lattice has two 
interpenetrating Bravais sublattices (FCC)
colored in blue (light)
and red (dark), respectively.
(a) The pairing amplitude
and 
(b) the second neighbour hopping 
are schematically shown,
where bonds colored by red (dark) and blue (light) 
have pairing potentials or second neighbour hopping amplitudes
which differ by a sign.
         }
\end{center}
\end{figure}\

Using this classification scheme which
was originally introduced in the context of disordered systems~\cite{Zirnbauer96}, 
it was shown in Ref.~\cite{Schnyder08} that, besides
the 3D $\mathbb{Z}_2$ topological insulator, there are precisely four more
3D topological quantum states.  Among these there is one which is particularly interesting from the point of view of  possible experimental realizations. It is called 
the topological superconductor in symmetry class CI
in the terminology of Ref.~\cite{Schnyder08} and
can be realized in time-reversal invariant singlet BCS superconductors. 
While the bulk is fully gapped 
in this topological quantum state, 
there are gapless robust Dirac states at the two-dimensional boundary.
The CI topological superconductor is unique among 3D topological quantum states in that it does not break $\mathrm{SU(2)}$ spin rotation symmetry,
and therefore supports the transport 
of spin through gapless surface modes.
The different topological phases of  the CI topological superconductor can be characterized by an 
integer winding number $\nu$, that takes on even values only, 
and which can be interpreted as the number of species of gapless surface Dirac fermions. 

In this Letter, we construct a lattice BCS Hamiltonian 
which realizes both the topologically trivial  and the nontrivial 
phases of the CI topological superconductor.
We compute 
the winding number and demonstrate the existence
of gapless two-dimensional Dirac fermions at the boundary.
The model we consider is defined on the diamond lattice with
(the lattice analogue of) 
{\it spin singlet} $d$-wave pairing, 
and has the form of a $4 \times 4$ Bogoliubov-de Gennes  (BdG) Hamiltonian.
In reciprocal space the noninteracting Hamiltonian reads
$H
=
\sum_{\bm{k}}
\Psi^{\dag}_{\bm{k}}
\mathcal{H}( \bm{k} )
\Psi^{\ }_{\bm{k}}
$,
with
$
\Psi^{\ }_{\bm{k}} =
(
 a^{\ }_{\bm{k} \uparrow}, b^{\ }_{\bm{k} \uparrow} , a^{\dag}_{-\bm{k} \downarrow}, b^{\dag}_{-\bm{k} \downarrow}
 )^T
$,
where
$a_{\bm{k},\alpha}$ and
$b_{\bm{k},\alpha}$  represent 
the electron annihilation operators with spin $\alpha$
and momentum $\boldsymbol{k}$
on sublattice $A$ and $B$ of the diamond lattice, respectively,
and 
\begin{eqnarray}
\mathcal{H}(\bm{k} )
\!\!&=&\!\!
\left(
\begin{array}{cccc}
\Theta_{\bm{k}} & \Phi^{\ }_{\bm{k}} & \Delta^{\ }_{\bm{k}} & 0 \\
\Phi^*_{\bm{k}} & -\Theta_{\bm{k}} & 0 &\Delta^{\ }_{\bm{k}} \\
\Delta^*_{\bm{k}} & 0 & -\Theta_{\bm{k}} & -\Phi^*_{-\bm{k}} \\
0 & \Delta^*_{\bm{k}} & -\Phi^{\ }_{-\bm{k}} & \Theta_{\bm{k}}
\end{array}
\right).
\label{eq: BdG Hamiltonian}
\end{eqnarray}
Here, the nearest neighbor hopping term is given by
$\Phi_{\bm{k}} =
\sum_{i=1}^4 t_i e^{{i} \bm{k} \cdot \bm{s}_i}$,
the next nearest neighbor hopping term is
$\Theta_{\bm{k}}
=
\sum_{i\neq j}
t'_{ij}
 e^{{i} \bm{k} \cdot ( \bm{s}_i- \bm{s}_j) }
+\mu_s$,
and the pairing potential is
$\Delta_{\bm{k}}
=
\sum_{i\neq j}
\Delta_{ij}
 e^{{i} \bm{k} \cdot ( \bm{s}_i- \bm{s}_j) }
$,
where $\bm{s}_{i=1\cdots 4}$ denotes the four first neighbor bond vectors.
The nearest and second nearest neighbor hopping amplitudes
are parametrized by the vector
$t_{l}$ and the symmetric matrix $t'_{ij}$, respectively.
Similarly, the symmetric matrix $\Delta_{ij}$ denotes
the singlet BCS pairing order parameter, whereas 
$\mu_s$ is the staggered chemical potential.

From Eq.~(\ref{eq: BdG Hamiltonian}), the energy eigenvalues 
$E^{\pm}_{\bm{k} } =
\pm \sqrt{ | \Phi_{\bm{k}} |^2 + | \Theta_{\bm{k}} |^2 + | \Delta_{\bm{k}} |^2 }  $
are readily obtained, exhibiting 
a two-fold degeneracy for each $\bm{k}$.
The symmetry operation that 
realizes particle-hole symmetry (PHS)
for a singlet pairing BdG Hamiltonian  is 
given by \cite{Schnyder08},
\begin{subequations} \label{eq: Class CI symmetries}
\begin{eqnarray}  \label{eq: Class CI PHS}
r_y \mathcal{H}^T(-\bm{k} ) r_y
= -\mathcal{H}( \bm{k} ) ,
\end{eqnarray}
where $r_y$ is the second Pauli matrix acting on the 
particle-hole space.
Hamiltonian (\ref{eq: BdG Hamiltonian})
automatically satisfies symmetry property (\ref{eq: Class CI PHS}), 
as can be checked easily.
If furthermore
time-reversal symmetry (TRS)
is present,
$\mathcal{H}(\bm{k} )$
obeys 
\begin{eqnarray} \label{eq: Class CI TRS}
\mathcal{H}^*(-\bm{k} ) = \mathcal{H}( \bm{k} ),
\end{eqnarray}
\end{subequations}
which is the case, if
the pairing amplitudes $\Delta_{ij}$ are all purely real.
The discrete symmetry constraints
(\ref{eq: Class CI PHS})
and (\ref{eq: Class CI TRS})
define the CI symmetry class 
in the Altland-Zirnbauer classification \cite{Zirnbauer96,Schnyder08}.
It is important to note, that an arbitrary Hamiltonian 
belonging to symmetry class CI
can be brought into   
block off-diagonal form.
This is achieved by means of a unitary transformation
which rotates
the $r_{\mu}$ matrices such that
$( r_x, r_y, r_z) \to ( r_x, -r_z, r_y)$.
Under this rotation $\mathcal{H}(\bm{k})$, Eq.~(\ref{eq: BdG Hamiltonian}), transforms  into 
\begin{subequations} \label{eq: off Ham}
\begin{eqnarray} \label{eq: off-diagonal}
\mathcal{H}( \bm{k} )
\to 
\left(
\begin{array}{cc}
0 & D( \bm{k} ) \\
D^{\dag}( \bm{k} ) & 0 
\end{array}
\right),
\end{eqnarray}
where the upper right block  is given by
\begin{eqnarray} \label{D operator}
D( \bm{k} )
=
\left(
\begin{array}{cc}
\Delta^{\ }_{\bm{k}} -{i}\Theta_{\bm{k}}  & -{i}\Phi_{\bm{k}} \\
-{i}\Phi^*_{\bm{k}} & \Delta^{\ }_{\bm{k}}+{i}\Theta_{\bm{k}} \\
\end{array}
\right) ,
\end{eqnarray}
\end{subequations}
which satisfies $D^T(-\bm{k})=D(\bm{k})$,
since
$\Delta_{-\bm{k}}=\Delta_{\bm{k}}$
and
$\Phi^*_{-\bm{k} }=\Phi_{\bm{k}}$.

In order to define a topological invariant for the CI topological insulator, we
need to introduce, following Ref.~\cite{Schnyder08}, 
the projection operator $Q(\bm{k})$
\begin{eqnarray}
Q(\bm{k} ) \!\!&=&\!\!
\left(
\begin{array}{cc}
0 & q(\bm{k} ) \\
q^{\dag}( \bm{k} ) & 0
\end{array}
\right),
\quad
q( \bm{k} )=
\frac{-D(\bm{k} )}{E^+(\bm{k})} ,
\end{eqnarray}
where, as a consequence of TRI and PHS, 
$ q^T(-\bm{k} )=q( \bm{k} )$.
The block off-diagonal form of $\mathcal{H} ( \bm{k} )$, and hence of 
$Q ( \bm{k} )$, is essential to uncover the 
topological structure
of the space of all possible quantum ground states
in class CI.  It allows us to introduce 
a topological invariant that classifies maps
from the Brillouin zone (BZ) 
into the space of  the projection operators $Q(\bf k)$.
This invariant is a winding number  defined 
in Ref.~\cite{Schnyder08} through
\begin{equation} \label{winding no}
\nu[q]
=
\int_{ }
\frac{d^3 k\,}{24\pi^2}
\epsilon^{\mu\nu\rho}\,
\mathrm{tr}\left[
\left( q^{-1}\partial_{\mu}q  \right)
\left( q^{-1}\partial_{\nu}q  \right)
\left( q^{-1}\partial_{\rho}q \right)
\right],
\end{equation}
where the integral is over the 3D BZ~\cite{footnote1}.
Due to the class CI constraint, $q^T(-\bm{k})=q( \bm{k} )$,
$\nu$ can take on only even integer values.
(In the absence of any constraint, i.e., for symmetry class AIII \cite{Schnyder08},
the winding number can be an arbitrary integer.)
The winding number changes only when the quantum system undergoes
a quantum phase transition, which is accompanied by the closing of the bulk gap.

We now turn to a more detailed specification
of our lattice model.
When $t_l=t$ 
for all $l=1,\ldots, 4$
\begin{align}
\Phi_{\bm{k} }
&=
2t 
\left[
e^{+{i} \frac{k_z}{4}}
\cos\frac{k_x+k_y}{4}
+
e^{-{i} \frac{k_z}{4}}
\cos\frac{k_x-k_y}{4}
\right].
\label{eq: nn hop}
\end{align}
The pairing potential we consider is 
``$d_{3z^2-r^2}$''-like, 
where the pairing amplitude on bonds
within the $x-y$ plane differs in sign
form the out-of-plane pairing amplitudes
[see Fig.\ \ref{fig: diamond lattice}(a)] 
\begin{eqnarray} \label{eq: d-wave pairing}
\Delta_{\bm{k}}
\!\!&=&\!\!
4 \Delta
\Big[
\cos \frac{k_x}{2} 
\cos \frac{k_y}{2}
-
\cos \frac{k_y}{2} 
\cos \frac{k_z}{2}
-
\cos \frac{k_z}{2} 
\cos \frac{k_x}{2}
\Big].
\nonumber\\
\end{eqnarray}
For definitiveness we will take $t$ and $\Delta$ positive
throughout the paper.
The BCS dispersion given by $\Phi_{\bm{k}}$ and
$\Delta_{\bm{k}}$ has four point nodes (Dirac points),
at which the bands are four-fold degenerate
\begin{eqnarray} \label{nodel points}
K_{1,\pm} =
2\pi \left(
\begin{array}{ccc}
\pm 1/3, & 1, & 0
\end{array}
\right),\,
K_{2,\pm} =
2\pi
\left(
\begin{array}{ccc}
1,& \pm 1/3, & 0
\end{array}
\right).
\nonumber
\end{eqnarray}
Note that 
PHS and TRI
relate  $K_{a,+}$ to $K_{a,-}$ ($a=1,2$), 
as $K_{a, -}$ is identified to $-K_{a, +}$ 
via a translation by  reciprocal lattice vectors. 
The degeneracy at the point nodes can be lifted by a finite staggered sublattice potential
or by second neighbor hopping amplitudes, which
are parametrized by the symmetric matrix  $t'_{ij}$.
In order to open up a bulk gap, we choose
$
t'_{13}
=
-t'_{14}
=+t'_{24}
=-t'_{23}
\equiv 
 t' \neq 0,
$
while $t'_{12}=t'_{34}=0$,
in which case the second neighbor 
hopping term simplifies to
[see Fig.\ \ref{fig: diamond lattice}(a)]
\begin{eqnarray}
\Theta_{\bm{k}} 
\!\!&=&\!\!
4t'
\cos \frac{k_z}{2}
\left(
\cos \frac{k_y}{2}-\cos \frac{k_x}{2}
\right)
+\mu_s,
\label{eq: nnn hop}
\end{eqnarray}
where we have also included a staggered chemical potential.

\begin{figure}
\includegraphics[width=0.42\textwidth]{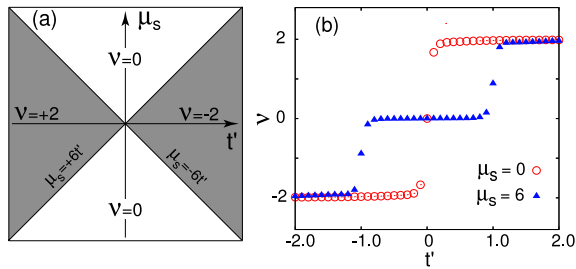}
\caption{
\label{fig: phase diagram}
(Color online) 
(a) Phase diagram as a function of second neighbor hopping
$t'$ and staggered chemical potential $\mu_s$. (b)
Numerical evaluation 
of the winding number.
}
\end{figure}

The BdG Hamiltonian (\ref{eq: BdG Hamiltonian}) has four energy bands,
two of which are occupied. With the choice for the hopping parameters
and gap amplitudes given by 
Eqs.\ (\ref{eq: nn hop}-\ref{eq: nnn hop}) the spectrum has 
a bulk energy gap, except when 
$\pm 6 t' = \mu_s$, 
i.e., when the system undergoes
a quantum phase transition between 
different topological phases.
The winding number $\nu$, Eq.~(\ref{winding no}),
for this model can be computed numerically,
by discretizing the integral over the BZ.
The numerical evaluation of $\nu$ as a function
of $\mu_s$ and $t'$ is shown in Fig.\ \ref{fig: phase diagram}(b), 
where we set $t = 4$, and $\Delta = 2$.
With increasing number of grid points in the BZ
the integral~(\ref{winding no}) converges rapidly 
to an even integer value. In this way we 
obtain the phase diagram shown in
Fig.\ \ref{fig: phase diagram}(a),
which contains four distinct gapped phases.
The nontrivial phases ($\nu=\pm 2$)
occur if $| \mu_s | < | 6 t' |$.
Inclusion of finite (but small) second neighbor terms
$t'_{12}$ and $t'_{34}$ shifts the phase boundaries,
but does not change the topology of the phase diagram.

When $\mu_s$ and $t'$ are small compared to $t$
[see  Fig.\ \ref{fig: phase diagram}(a)],
an effective low-energy continuum description
can be derived 
by expanding Hamiltonian~(\ref{eq: off Ham}) around the four nodal points $K_{1/2,\pm}$. 
Rescaling momenta as 
$t k_z/2 \to k_z$,
$t \sqrt{3}k_y/2 \to k_y$,
and 
$2 \sqrt{3} \Delta k_x \to k_x$,
and performing a unitray transformation, 
we find that the low-energy expansion   around the node $K_{1\pm}$  
of the off-diagonal block in Eq.~(\ref{eq: off-diagonal})
is given by
\begin{eqnarray} \label{eq: rotated D1}
D(\bm{q})
=
\beta  {i}\sigma_y
\left[
\tilde{\bm{q} } \cdot \boldsymbol{\alpha} 
-{i}
(\mu_s + 6t') \gamma^5
\right] ,
\end{eqnarray}
where $\tilde{\bm{q}}=(q_x,-q_y,-q_z)$
denotes the deviation of the momentum from $K_{1\pm}$.
Eq.~(\ref{eq: rotated D1}) is identical to the class CI topological
Dirac superconductor constructed in Ref.~\cite{Schnyder08}. 
Here, we have introduced the five
gamma matrices $\alpha_{\mu}$, $\beta$, and
$\gamma^5$, which are in the Dirac representation
given by 
$\alpha_{\mu}   =
\sigma_{\mu}\otimes \tau_x$,
$\beta 
= 
\tau_z$,
and
$\gamma^5
= 
\tau_x$,
with $\tau_{x,y,z}$ being another set of Pauli matrices.
Consequently, the winding number within the continuum description
for the nodes 
$K_{1,\pm}$ is
\begin{eqnarray}
\nu_{1} = \frac{1}{2}
\frac{(\mu_{s}+6 t')}{|\mu_s+6 t'|}
\times 2,
\end{eqnarray}
where the prefactor $1/2$, which is an articfact of the low-energy continuum approximation,  
will be altered once 
information about the structure of Bloch wavefunctions
at high energy (located away from the Dirac points) is 
supplemented.
The low-energy Dirac Hamltonian for the nodes 
$K_{2,\pm}$ is
related by symmetry to the result
for the nodes $K_{1\pm}$ 
by simultaneously interchanging $k_x$ with $k_y$
and replacing the mass term $\mu_s + 6t'$
with $\mu_s - 6t'$,
\begin{eqnarray}
D(\bm{q})
=
\beta  {i}\sigma_y
\left[
\tilde{\bm{q}} \cdot \boldsymbol{\alpha} -{i}
(\mu_s - 6t') \gamma^5
\right],
\end{eqnarray}
where $\tilde{\bm{q}}=(q_y,-q_x,-q_z)$.
Similarly, the winding number within
the continuum description for the nodes $K_{2,\pm}$ is
\begin{eqnarray}
\nu_{2} = -\frac{1}{2}
\frac{(\mu_{s}-6t')}{|\mu_s-6t'|}
\times 2. 
\end{eqnarray}
Interestingly,
the winding number obtained from the continnum description,
$\nu = \nu_{1} + \nu_{2}$,
reproduces the phase diagram
of the lattice model correctly. 
 
\begin{figure}
  \begin{center}
\includegraphics[width=0.42\textwidth]{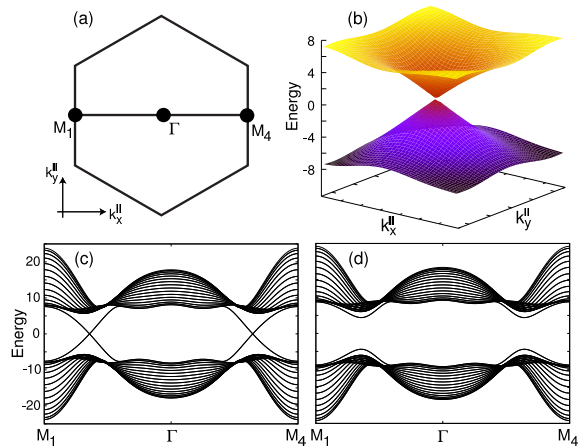} 
\caption{
\label{fig: surf. Bz}
(Color online) 
(a) Surface Brillouin zone 
for a slab parallel to the (111) surface.
Panels (c) and (d) show the
two-dimensional band structure
for the model defined by  
Eqs.\ (\ref{eq: BdG Hamiltonian}, \ref{eq: nn hop}-\ref{eq: nnn hop})
terminated by a (111) surface, with 
the momentum being varied
along the line indicated in panel (a).
For $(t',\mu)=(1.2,0)$ [panel (c)]
there are surface states which cross the 
bulk energy gap, thereby realizing a surface
Dirac fermion state as shown in (b). In the
trivial phase, for $(t',\mu)=(0,4.4)$ [panel (d)],
there are no surface states crossing the energy gap.
}
\end{center}
\end{figure}

A physical consequence of
the non-zero winding number $\nu$ is the appearance
of zero-energy surface Andreev bound states with Dirac dispersion.
To study these surface states we solve
model (\ref{eq: BdG Hamiltonian}) in a slab geometry,
and compute the energy bands for a slab parallel
to the (111) surface, both in the trivial and 
nontrivial phases.
As shown in Fig.~\ref{fig: surf. Bz},
where we set $t=4$ and $\Delta=2$, there are, in addition
to the bulk states, surface Dirac states which cross the band gap. 
When the bulk topological invariant is  $\nu=2$ [Fig.~\ref{fig: surf. Bz}(c)]
there are two surface Dirac states, whereas there is no such state
when  the bulk topological invariant is  $\nu=0$ [Fig.~\ref{fig: surf. Bz}(d)].
Hence, the total number of Dirac states/cones $N_f=2$
is consistent with the bulk characteristics $\nu=2$.
$N_f=2$ is the minimal number of Dirac cones required by
class CI symmetries. 
Note, however, that in  any two-dimensional lattice model
satisfying the CI symmetries, the possible number of Dirac 
cones is an integer 
multiple of $2N_f$ because of
a no-go theorem analogous to the fermion
doubling theorem of Ref.\ \cite{nielsen1981}.
Here,  the fermion doubling 
is avoided since the two-dimensional
system is realized as a boundary of a 3D bulk.


The surface Dirac fermion modes cannot be gapped  
by any deformation of the Hamiltonian respecting 
time-reversal and particle-hole symmetry. 
Indeed,  this is so because any perturbation
respecting the class CI symmetries
takes the form of 
an SU(2) gauge field,
which perturbs the 
surface Dirac fermions
in the following way
\begin{align}
\mathcal{H}
&=
(k_x + \boldsymbol{a}_x \cdot \boldsymbol{\sigma})\tau_x 
+
(k_x + \boldsymbol{a}_y \cdot \boldsymbol{\sigma})\tau_y,
\label{eq: surface Dirac}
\end{align}
where $a^{a=1,2,3}_{\mu=x,y}\in \mathbb{R}$.
The gapless nature of
this four-component Dirac fermion is
stable against arbitrary values of
the six real parameters $a^{a}_{\mu}$,
i.e., the non-Abelian gauge potential 
shifts the location of the Dirac node, but does not lead
to a gap.
With the inclusion of randomness, small relative to the bulk energy gap,
the surface Dirac fermion mode realizes the 
random SU(2) gauge potential model
discussed in 
Ref.\ \cite{Tsvelik95}.
The random SU(2) gauge potential, known not to be able to
localize the Dirac fermions, renormalizes to an
exactly solved strong coupling
renormalization group (RG) fixed point at long distances.
Under the influence of disorder, the (tunneling) density of states
$\rho(\epsilon)$ changes from a linear dependence to  
$\rho(\epsilon) \sim | \epsilon|^{\mu}$, 
with the scaling exponent $\mu=1/7$ \cite{Tsvelik95}
at this fixed point.

Within the non-linear sigma model approach  based on $n$ fermionic
replicas ($n\to 0$),
the Anderson localization physics
at the surface of the 3D class CI topological superconductor
in the presence of disorder is described by the
principal chiral model on $\mathrm{Sp}(2n)$, 
supplemented by a Wess-Zumino-Witten (WZW) term, at `level-one'.
(When supersymmetric disorder averaging is used, this is~\cite{SUSYDisorder} the
WZW model on $\mathrm{Osp}(2|2)$ at level $-\nu$~\cite{GeneralWZW}.)
We emphasise that the WZW term cannot appear for
any two-dimensional disordered systems on a lattice,
such as disordered $d_{x^2-y^2}$-wave superconductors
on the square lattice.
Indeed, WZW terms always cancel when contributions from 
different cones are added up. 
Thus, the delocalized feature (non-renormalization of
the conductivity) on a 2D lattice is observed, at best, 
as a crossover behavior when the inter-cone scatterings
are weak.


As it stands, the lattice model we considered is presumably not directly 
connected to any specific system occurring in nature;
but it may give insight into the properties of real materials that exhibit three-dimensional
topological phases which preserve both time-reversal and spin-rotation symmetry. 
Promising candidates for the CI topological quantum state
might be found among some of the unconventional superconductors
of heavy fermion systems.
There are various ways
how the topologically protected surface states
in such compounds could be detected experimentally.
First of all, 
because of spin rotation symmetry,
the surface conductivity for the spin current is a well-defined quantity,
which is unchanged by symmetry preserving perturbations, including disorder.
Secondly, the density of states of the surface states
can be probed via 
tunneling experiments.
Finally, we argue that by breaking TRS locally at the surface,
while keeping the spin-rotation symmetry intact,
one can realize the so-called `{\it spin quantum Hall effect} (SQHE)'\cite{senthil99}
in symmetry class C
(unrelated to the `quantum spin Hall insulators' mentioned earlier)
at the surface of the CI topological superconductor.
This can be seen as follows.  Broken 
TRS (without breaking SU(2) symmetry) allows for the appearance of 
four additional perturbing potentials
in Eq. (\ref{eq: surface Dirac}), 
describing
the surface modes.
One can check that only one of theses additional potentials gives rise to a gap,
and hence to a (2D)
non-zero Chern integer $n=\pm 1$; this is one half of
the allowed value of the Chern number 
in any 2D system exhibiting the SQHE \cite{senthil99}.
This situation is completely analogous to the half-integer surface QHE
of the 3D $\mathbb{Z}_2$ topological insulator \cite{Fu06_3Da,Qi2008, Essin08}.

In the 3D $\mathbb{Z}_2$ topological insulator,
this topological magneto-electric effect can be described by 
the effective field theory whose action is given by
(3+1) QED supplemented with $\theta=\pi$ term  \cite{Qi2008, Essin08}.
Since spin is a good quantum number in a singlet superconductor,
it is possible to describe its spin transport
in terms of an external SU(2) gauge field.
Indeed, the effective field theory of the SQHE in TRS-breaking 2D singlet superconductors 
is known to be the SU(2) Chern-Simons theory at an integer level \cite{ReadGreen}.
Then, following the same reasoning
as in the case of the 3D $\mathbb{Z}_2$ topological insulators,
we propose that 
the effective field theory describing 
class CI topological superconductor is the (3+1)D SU(2) Yang-Mills theory
argumented with the theta term
\begin{eqnarray}
\mathcal{L}
=
\frac{\theta}{32\pi^2}
\epsilon^{\mu\nu\kappa\lambda}
\mathrm{tr}\,\left(
F_{\mu\nu} F_{\kappa\lambda}
\right),
\end{eqnarray}
where $F_{\mu\nu}$ is the $\mathrm{SU}(2)$ field strength
and $\theta=\pi$.

\acknowledgments

We are grateful to the KITP and all the participants 
of the quantum spin hall Mini-program.
This work has been supported in part by the National Science Foundation
(NSF) under Grant Numbers PHY05-51164 (S.R., A.S.)
and DMR-0706140 (A.W.W.L.).
S.R.\ thanks the Center for Condensed Matter Theory
at University of California, Berkeley for its support.



\begin{thebibliography}{99}

\bibitem{KaneMele}
C.\ L.\ Kane and E.\ J.\ Mele,
Phys.\ Rev.\ Lett.\ \textbf{95}, 146802 (2005);
\textbf{95}, 226801 (2005).

\bibitem{bernevig06}
B.\ A.\ Bernevig, T.\ L.\ Hughes, and S.-C.~Zhang, 
Science \textbf{314}, 1757 (2006).

\bibitem{moore07}
J.\ E.\ Moore and L.\ Balents,
Phys.\ Rev.\ B \textbf{75}, 121306(R) (2007).

\bibitem{Fu06_3Da}
L.\ Fu, C.\ L.\ Kane, and E.\ J.\ Mele,
Phys.\ Rev.\ Lett.\ \textbf{98}, 106803 (2007).

\bibitem{Fu06_3Db}
L.\ Fu and C.\ L.\ Kane,
Phys.\ Rev.\ B \textbf{76}, 045302 (2007).

\bibitem{Schnyder08}
Andreas P.\ Schnyder,
Shinsei Ryu,
Akira Furusaki,
Andreas W.\ W.\ Ludwig,
Phys.\ Rev.\ B \textbf{78}, 195125 (2008);
see also
 ``Anderson Localization at Boundaries: 
Classification of Topological Insulators and Superconductors'',
http://landau100.itp.ac.ru/Talks/ludwig.pdf.

\bibitem{konig07}
M.\ K\"onig \textit{et al.},
Science \textbf{318}, 766 (2007).

\bibitem{hasan07}
D.\ Hsieh, 
D.\ Qian, 
L.\ Wray, 
Y.\ Xia, 
Y.\ Hor, 
R.\ J.\ Cava, 
and M.\ Z.\ Hasan, 
Nature \textbf{452}, 970 (2008).

\bibitem{Kitaev08}
The same conclusion has recently been reached 
from K-theory by A.\ Yu.\ Kitaev,
``Periodic table for topological insulators and superconductors'',
http://landau100.itp.ac.ru/Talks/kitaev.pdf.

\bibitem{Zirnbauer96}
M.\ R.\ Zirnbauer,
J.\ Math.\ Phys.\ \textbf{37}, 4986 (1996);
A.\ Altland and M.\ R.\ Zirnbauer,
Phys.\ Rev.\ B \textbf{55}, 1142 (1997);
P. Heinzner, A. Huck Leberry, and M. R. Zirnbauer,
Commun.\ Math.\ Phys. \textbf{257}, 725 (2005).

\bibitem{footnote1}
A topological invariant of a similar type was discussed
in the context of the
B phase in ${ }^{3}$He by M.\ M.\ Salomaa and G.\ E.\ Volovik,
Phys.\ Rev.\ B \textbf{37}, 9298 (1988).

\bibitem{nielsen1981}
H.\ B.\ Nielsen and M.\ Ninomiya,
Nucl.\ Phys.\ B \textbf{185}, 20 (1981).

\bibitem{Tsvelik95}
A.\ M.\ Tsvelik,
Phys.\ Rev.\ B \textbf{51}, 9449 (1995).

\bibitem{SUSYDisorder} 
M. J. Bhaseen \textit{et al.}, Nucl. Phys. B \textbf{618}, 465 (2001); A. W. W. Ludwig, arXiv:cond-mat/0012189.

\bibitem{GeneralWZW}
More generally, on surfaces of 3D topological insulators
in classes AIII and DIII of Ref. \cite{Schnyder08} the WZW model
on $Gl(1|1)_{\nu}$ and $Osp(2|2)_{\nu}$ appears in the presence of disorder,
respectively,
where $\nu$ is the topological winding number in the bulk. This follows 
from Ref.~\cite{Schnyder08} and, e.g.,  
A. LeClair and D. J. Robinson, J. Phys. A \textbf{41}, 452002 (2008).

\bibitem{senthil99}
T.\ Senthil, J.\ B.\ Marston, and M.\ P.\ A.\ Fisher, Phys.\ Rev.\ B \textbf{60}, 
4245 (1999); I.\ A.\ Gruzberg, A.\ W.\ W.\ Ludwig, and N.\ Read, 
Phys.\ Rev.\ Lett.\ \textbf{82}, 4524 (1999).

\bibitem{Qi2008}
Xiao-Liang Qi, Taylor Hughes, Shou-Cheng Zhang,
\texttt{arXiv:0802.3537}

\bibitem{Essin08}
A.\ M.\ Essin, J.\ E.\ Moore, and D.\ Vanderbilt,
\texttt{arXiv:0810.2998}.

\bibitem{ReadGreen}
N.\ Read and D.\ Green,
Phys.\ Rev.\ B \textbf{61} 10267, (2000).

\end{thebibliography}
\end{document}